\def\year{2019}\relax
\documentclass[letterpaper]{article} 
\usepackage{aaai19}  
\usepackage{times}  
\usepackage{helvet}  
\usepackage{courier}  
\usepackage{url}  
\usepackage{graphicx}  
\usepackage{booktabs} 
\usepackage{xcolor}
\usepackage{color}
\usepackage{colortbl}
\usepackage{soul}
\usepackage{amsfonts}
\usepackage[small]{caption}
\usepackage{balance} 
\usepackage[linesnumbered,ruled,vlined]{algorithm2e}
\usepackage[flushleft]{threeparttable}
\usepackage{tablefootnote}
\usepackage{enumitem}
\usepackage{graphics}
\usepackage{subcaption}
\usepackage{amsmath}
\usepackage{comment}
\usepackage{float}
\usepackage{varioref}
\usepackage{stfloats}
\usepackage{pifont}
\newcommand*{\affaddr}[1]{#1} 

\newcommand*{\email}[1]{\texttt{#1}}
\usepackage{array}
\usepackage{float}
\newcolumntype{P}[1]{>{\centering\arraybackslash}p{#1}}
\newcolumntype{M}[1]{>{\centering\arraybackslash}m{#1}}

\begin{document}
%
\title{EndCold: An End-to-End Framework for Cold Question Routing in Community Question Answering Sites}
\author{%
\small Jiankai Sun,
Jie Zhao, Huan Sun, and Srinivasan Parthasarathy\\
\affaddr{\small  Department of Computer Science and Engineering, The Ohio State University, USA}\\
\email{\footnotesize \{sun.1306,zhao.1359,sun.397\}.osu.edu,srini@cse.ohio-state.edu}\\
}
\maketitle
\begin{abstract}
Routing newly posted questions (a.k.a cold questions) to potential answerers with the suitable expertise in Community Question Answering sites (CQAs) is an important and challenging task.
The existing methods either focus only on embedding the graph structural information and are less effective for newly posted questions, or adopt manually engineered feature vectors that are not as representative as the graph embedding methods.
Therefore, we propose to address the challenge of leveraging heterogeneous graph and textual information for cold question routing by designing an end-to-end framework that jointly learns CQA node embeddings and finds best answerers for cold questions.
We conducted extensive experiments to confirm the usefulness of incorporating the textual information from question tags and demonstrate that an end-2-end framework can achieve promising performances on routing newly posted questions asked by both existing users and newly registered users.
\end{abstract}

\section{Introduction}
\label{sec:introduction}

Community Question Answering services (CQAs) such as Stack Exchange and Yahoo! Answers are examples of social media sites, with their usage being examples of an important type of computer supported cooperative work in practice. In recent years, the usage of CQAs
has seen a dramatic increase in both the frequency of questions posted and general user activity.
This, in turn, has given rise to several interesting problems ranging from expertise estimation to question difficulty estimation, and from automated question routing to incentive mechanism design on such CQAs ~\cite{Yang2013CQArank,Fang2016QuestionAnswering,QDEE2018}. 

In this work, we focus on the problem of routing newly posted questions to suitable experts before answers are written (item cold-start). Usually, there are two types of questions in CQAs -- resolved (questions with answers) and newly posted questions (questions that have not received any answers). The newly posted questions may themselves be posted by new askers (such as newly registered users who have not asked a question earlier) or existing askers (such as users who have asked several questions previously).
We refer to these newly posted questions as {\em cold questions}. The majority of approaches have focused on evaluating content quality after the fact (after questions have been resolved) ~\cite{Yang2013CQArank}. Yet, as the CQAs continue to grow, routing the cold questions to matching experts before answers have been provided has become a critical problem. For example, in Stack Overflow, about $5.22$ million questions have not been answered \footnote{\url{https://stackexchange.com/sites#}}. 

Recently, the usage of graphs has seen an increase in solving the question routing problem. For example, Sun et al.~\cite{QDEE2018} proposed {\em QDEE} to leverage Expertise Gain Assumption (EGA) to build the competition graph and then applied social agony~\cite{Tatti2014,tatti2015} to infer graph hierarchy and assign each node a scalar value to represent where it stands in the competition graph. The assigned scalar value of question nodes are question difficulty scores, and the assigned scalar value of user nodes are user expertise scores. {\em QDEE} then routed cold questions to users with matching expertise based on the corresponding questions' difficulty level. 
It's worth mentioning that both question difficulty and user expertise follow the characteristic of asymmetric transitivity. For example, given a question $q_1$ is easier than $q_2$ and $q_2$ is easier than $q_3$, we can infer that $q_1$ is easier than $q_3$ conveniently. 
Sun et al. ~\cite{Sun2018ATP} proposed {\em ATP} to tackle the challenge of directed graph embedding with asymmetric transitivity preservation and then they leveraged the proposed embedding results to appropriately route and assign newly posted questions to users with the suitable expertise and interest in CQAs. However, a multiple pipeline including directed graph embedding generation and experts finding is required where each step has to be optimized separately. Another drawback is that {\em ATP} can only leverage graph structure information, while other information (such as textual information) that does not follow the asymmetric transitivity can not be incorporated. It limits {\em ATP} in inferring the embedding of cold questions asked by new askers who have no activities in CQAs and hence {\em ATP} cannot work well for routing such questions to suitable answerers.

 In Stack Exchange sites, voting is central for providing quality questions and answers \footnote{\url{https://stackoverflow.com/help/why-vote}}. 
 The more that people vote on a post, the more certain future visitors can be confident of the quality of information contained within the post. Voting indicates a CQA community's long-term review for a given user's expertise level under a specific topic.   ColdRoute~\cite{ColdRoute} views the problem of identification of best answerers as a regression problem (finding the answerers who have the highest predicted voting scores). 
The input features of {\em ColdRoute} are users' past asking and answering activities in CQAs, and {\em ColdRoute} can leverage Factorization Machines (FMs)~\cite{Rendle2010FM,Rendle2012libFM} to model pairwise interactions among different objects (questions, askers, answerers, and question tags). While in theory FMs can model high-order interactions, in practice it only models order-2 feature interactions due to the high complexity~\cite{ijcai2017deepfm}.    

Therefore, we propose to solve the cold question routing problem by marrying the merits of the above two types of methods. On one hand, we use graph embedding techniques to encode useful high-order graph structure information, and on the other hand, we enable the incorporation of heterogeneous information (e.g. textual) in the node embedding as well. Specifically, \textit{users} (including both askers and answerers), \textit{questions}, and \textit{question tags} are represented as three different types of nodes, which are connected by their historical interactions into an undirected CQA graph. 
We then employ a Graph Convolutional Network (GCN)~\cite{kipf2016semi} to generate deeper and neighbor-aware node representations in the undirected graph.
 Our proposed method allows more flexible node interactions and can enrich the question/user node embeddings with the textual information from question tags. And question tags can bridge the gap between cold questions and exiting nodes in the CQA graph. 
This overall end-to-end framework can be optimized directly with the cold question routing objective, which enables the learning of task-specific embeddings for heterogeneous nodes in the CQA graph, and a better cold question routing performance.

In summary, we made the following contributions:
\begin{itemize}[leftmargin=0.6cm]
    \item Encode users' past asking and answering activities and textual information by using an undirected heterogeneous graph, of which nodes are questions, tags, and users (askers and answerers) and edges are interactions among them.
    \item Build an end-to-end framework which can route cold questions in CQAs effectively.
    \item Demonstrate the effectiveness of our proposed framework on various Stack Exchange sites.
\end{itemize}

\section{Related Work}
\label{sec:relatedWork}

Recently, Sun et al.~\cite{QDEE2018} observed that users typically gain expertise across multiple interactions with the CQA and tend to ask more difficult questions within the same domain over time, which is referred to as {\em Expertise Gain Assumption} (EGA). They leveraged EGA and added additional edges between questions asked by the same user to the previous competition graph~\cite{Liu2011,liu2013question,wang2014} to combat the sparseness problem. They then proposed {\em QDEE} to lever social agony~\cite{Tatti2014,tatti2015} to infer graph hierarchy and assign each node a scalar value to represent where it stands in the competition graph. The assigned scalar value of question nodes are question difficulty scores, and the assigned scalar value of user nodes are user expertise scores.
There is some work which can model the users' past asking and answering activities by using a graph and then some graph embedding methods~\cite{Fang2016QuestionAnswering,zhaoexpert2016,zhao2017community} are proposed to address the above limitation. 
The problem of graph embedding seeks to represent vertices of a graph in a low-dimensional vector space in which meaningful semantic, relational and structural information conveyed by the graph can be accurately captured~\cite{Ma2018WSDM}. 
Recently, one has seen a surge of interest in developing such methods including ones for learning such representations for directed graphs while preserving important properties such as asymmetric transitivity~\cite{Ou2016KDDAsymmetric,Sun2018ATP}. 
Question difficulty and user expertise follow the characteristic of asymmetric transitivity. For example, given a question $q_1$ is easier than $q_2$ and $q_2$ is easier than $q_3$, we can infer that $q_1$ is easier than $q_3$ easily. It happens to estimate user expertise too. We can infer that $u_1$ has more expertise than $u_3$ based on the fact that $u_1$ has more expertise than $u_2$ and $u_2$ has more expertise than $u_3$ in a specific domain. Sun et al. ~\cite{Sun2018ATP} proposed {\em ATP} to tackle the challenge of directed graph embedding with asymmetric transitivity preservation and then they leveraged the proposed embedding results to appropriately route and assign newly posted questions to users with the suitable expertise and interest in CQAs. However, a multiple pipeline including directed graph embedding generation and experts finding is required where each step has to be optimized separately. Another drawback is that {\em ATP} can only leverage graph structure information while cannot leverage textual information (such as question tags) since these textual information cannot be naturally leveraged to construct the input directed graph. It indicates that {\em ATP} has some limitation on inferring the embedding of cold questions asked by new askers who have no activities in CQAs and hence cannot work well for routing cold question asked by new askers to suitable answerers. 

Question difficulty and user expertise can vary in different topics. In Stack Exchange sites, users are required to use tags (a tag is a word or phrase) to describe the topic(s) of the question \footnote{\url{https://stackoverflow.com/help/tagging}}. Each question can be assigned multi-tags to represent its most relevant topics. 
Hence tags are important user-generated category information that achieves fine-grained and dynamic topic representation. Users who use a particular tag when posting questions or answers might prefer topic summaries most relevant to that tag~\cite{LabeledLDA2009}. In our experiments, the average number of tags per question is $2.82$ and $2.96$ in Stack Exchange site Apple and Physics respectively. Hence a solely scalar value to represent question difficulty level or user expertise is not thorough. By iteratively introducing questions tags and textual descriptions (question title and body), {\em ColdRoute}~\cite{ColdRoute} demonstrated that question tags play a more important role than question title and body in terms of experts finding. {\em ColdRoute} views the problem of identification of best answerers as finding the answerers who have the highest predicted voting scores. Given a cold question $q$ and a set of potential answerers $C_q$, {\em ColdRoute} predicts each candidate $u$'s voting score for $q$, where $u \in C_q$, and then selects the user who achieves the highest predicted voting score as the best answerer for $q$. 
In Stack Exchange sites, voting is central for providing quality questions and answers \footnote{\url{https://stackoverflow.com/help/why-vote}}. Voting up a post signals to the rest of the community that the post is interesting, well-researched, and useful. A highly voted post reflects the quality of the post -- which may be viewed by the future visitors. 
The more that people vote on a post, the more certain future visitors can be confident of the quality of information contained within the post. Voting indicates a CQA community's long-term review for a given user's expertise level under a specific topic. 
Each voting score is an integer, which is calculated based on the difference between the corresponding answer's up-votes and down-votes which are assigned to it by users who viewed the question or provided answers in the CQAs. The input features of {\em ColdRoute} are users' past asking and answering activities in CQAs, and {\em ColdRoute} can leverage Factorization Machines (FMs)~\cite{Rendle2010FM,Rendle2012libFM} to model the interactions among different objects (questions, askers, answerers, and question tags). 


\subsection{Problem Statement}
\label{sec:ProblemStatement}

Assume we are given four relational sets of data in terms of Questions $\mathcal{Q} = \langle q_1, q_2, \dots, q_n\rangle$, Askers $\mathcal{A} = \langle a_1, a_2, \dots, a_m\rangle$, Answerers $\mathcal{U} = \langle u_1, u_2, \dots, u_k\rangle$, and Question Tags $\mathcal{T} = \langle t_1, t_2, \dots, t_l\rangle$. For each question $q_i \in \mathcal{Q}$, we have a tuple of the form $\langle Asker_i,$ $Answerers_i$, $BestAnswerer_i$, $Tags_i$, and $Scores_i\rangle$, where $Asker_i \in \mathcal{A}$, $Answerers_i \subset \mathcal{U}$, $BestAnswerer_i \in \mathcal{U}$ , $Tags_i \subset \mathcal{T}$. Each voting $score \in Scores_i$ is an integer, which is calculated based on the difference between $Answerer_i$'s up-votes and down-votes which are assigned to it by users who viewed the question or provided answers for that in the CQA environment. Note that the $BestAnswerer$ for a question may not be specified by $Asker$. 

Given the preliminaries (above), in this work, we focus on the problem of routing newly posted questions to suitable experts before answers are written (item cold-start).
Each quadruple case $\langle q,u,a,t \rangle$, where $q \in \mathcal{Q}$, $u \in \mathcal{U}$, $a \in \mathcal{A}$, $t \subset \mathcal{T}$ has a voting score $y \in \mathbb{R}$, which is equal to the difference between times of up-voting and down-voting. Given the testing question set $\mathcal{Q}_t$, the predicted ranking list of all potential answerers for a test question $q^*$ is $R^{q^*}$ for all $q^* \in \mathcal{Q}_t$. The answerer who's ranked top $1$ among all candidates in $R^{q^*}$ for the corresponding target question $q^*$ will be selected as the best answerer for $q^*$. 

\section{Methodology}
\label{sec:methodolody}

\begin{figure}[ht]
    \centering
    \includegraphics[width=0.5\textwidth]{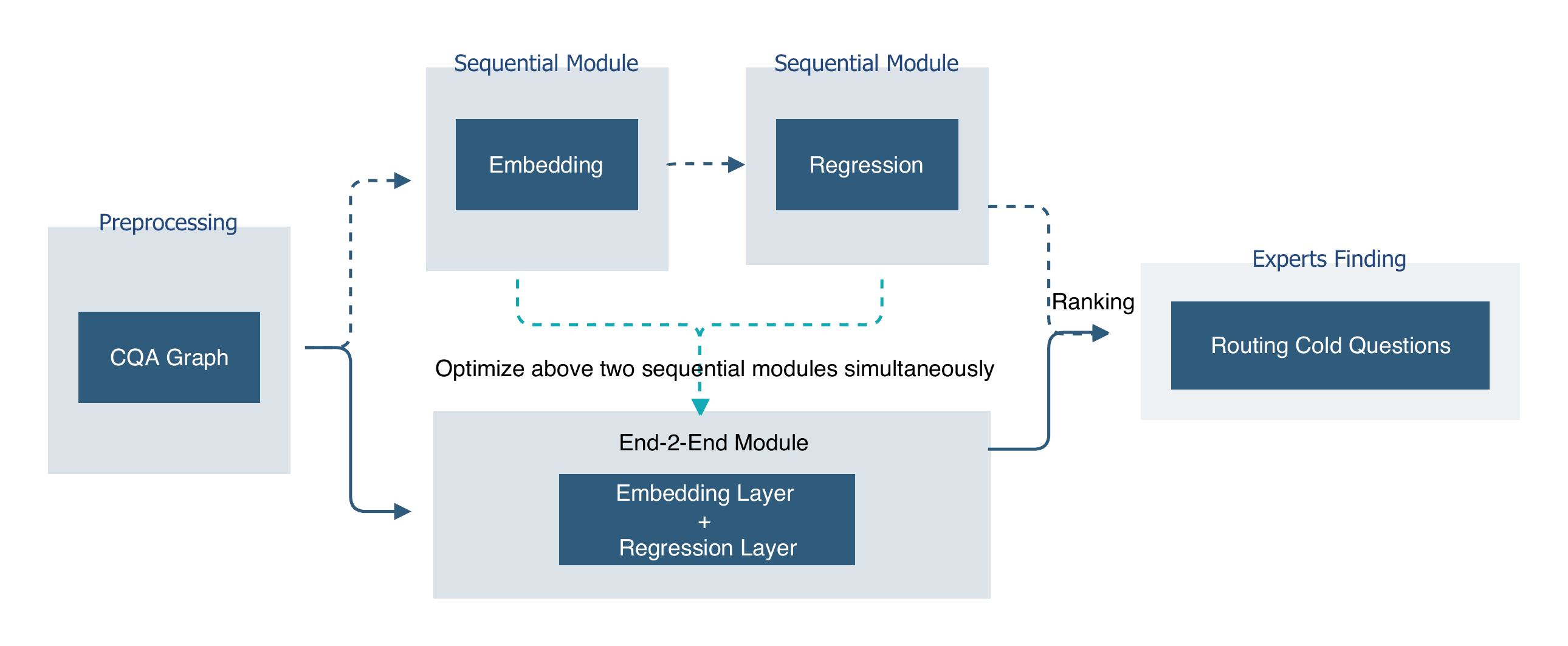}
    \caption{Illustration of the sequential and end-2-end framework for cold question routing in CQAs} 
    \label{fig:endcold_framework}
\end{figure}

In this section, we talk about the design of our proposed framework (both sequential and end-2-end). The key steps of the sequential and end-2-end model for routing newly posted questions are shown in Figure~\ref{fig:endcold_framework}. We refer the sequential model as {\em Seq} and the end-2-end model as {\em EndCold}. The key steps of {\em Seq} are:

\begin{enumerate}[leftmargin=0.4cm]
    \item Build the undirected heterogeneous CQA graph which incorporates users past asking and answering activities and their corresponding textual information in CQAs
    \item Apply suitable graph embedding methods to learn representations for each node (including question, user, and tag) in the corresponding input CQA graph
    \item Model the question routing problem as a regression problem, and concatenate the embeddings of question, asker, answerer, and tag as feature vectors and use corresponding voting scores as targets to train a suitable regression model $f$
    \item Apply $f$ to predict the voting score a candidate user can achieve on a given cold question $q^*$
    \item Routing $q^*$ to its corresponding predicted best answerer, identified by selecting the user who's predicted to achieve the highest voting score among all candidate answerers
\end{enumerate}

The biggest difference between our sequential model {\em Seq} and end-2-end framework {\em EndCold} is that {\em EndCold} optimizes the graph embedding module (step $2$ in {\em Seq}) and regression module (step $3$ in {\em Seq}) simultaneously as shown in Figure~\ref{fig:endcold_framework}. 

\subsection{Build the undirected CQA graph}

\begin{figure}[ht]
    \centering
    \includegraphics[width=0.45\textwidth]{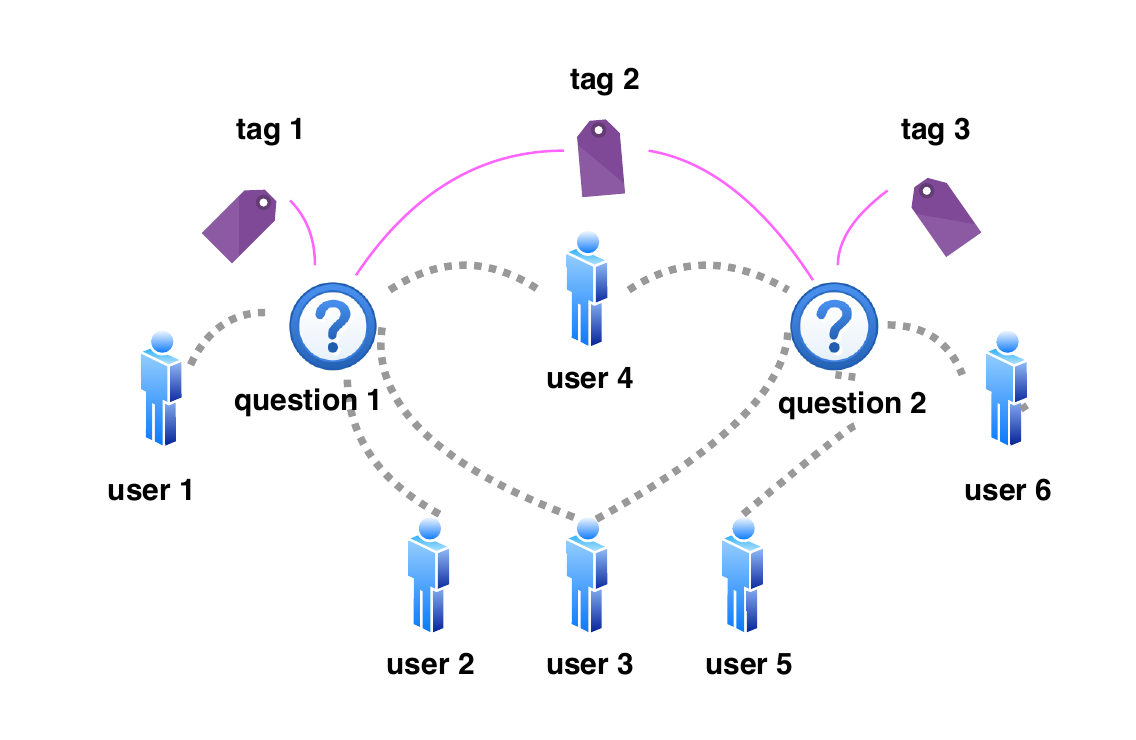}
    \caption{Illustration of the input CQA Graph with question tags: user $1$ asked a question $1$ which has tags $1$ and $2$. User $2$, $3$, and $4$ answered question $1$. User $4$ asked a question $2$ with tag $2$ and $3$. Question $2$ was answered by user $3$, $5$ and $6$.} 
    \label{fig:cqa_graph_illustration}
\end{figure}

Figure~\ref{fig:cqa_graph_illustration} illustrates how we build the corresponding heterogeneous CQA graph of which nodes are questions, tags, and users (including askers and answerers) and edges represent the interactions among questions, askers, answerers, and question tags. User $1$ asked a question $1$ which has tags $1$ and $2$. User $2$, $3$, and $4$ answered question $1$. User $4$ asked a question $2$ with tag $2$ and $3$. Question $2$ was answered by user $3$, $5$ and $6$. Edges are interactions among all questions, tags, askers and answerers. 

Both users' past asking and answering activities and their corresponding textual information (question tags) are incorporated in the above CQA graph, which can help us bridge the gap between cold question nodes and existing nodes in the CQA graph. Cold questions can be connected with the existing CQA graph via the common question tag nodes they have and existing asker node (not applicable for cold questions asked by new askers). 

\subsection{The Sequential Module for Cold Question Routing}
\label{sec:node2vec_un_at}

In this section, we talk about the sequential model named as {\em Seq-TA} (By default, {\em Seq} is referred to {\em Seq-TA} in our paper) which takes the undirected CQA Graph with question tags as the input and leverages any suitable graph embedding algorithms to generate embeddings for the question, user, and tag nodes. Take our experiments for example, we use node2vec~\cite{grovernode2vec} to generate corresponding embeddings\footnote{Other state-of-the-art undirected graph embedding methods can be applied to this module too}. Each answering thread between a question and an answerer can be represented as a quadruple of the target question, its asker, the corresponding answerer, and question tags. The corresponding quadruple case $\langle q,u,a,t \rangle \in \langle \mathcal{Q},\mathcal{U},\mathcal{A},\mathcal{T} \rangle$ can be represented as a concatenated feature vector $(\vec{q},\vec{u},\vec{a},\vec{t})$, where $\vec{q}$, $\vec{u}$, and $\vec{a}$ are the embedding vectors of question $q$, answerer $u$, and asker $a$ respectively. $\vec{t}$ is the average embedding of tags in set $t$. The voting score of $u$ achieved on $q$ is an integer. A regression model $f$ is trained based on the above feature vectors as inputs and corresponding voting scores as the targets. 

This design gives us the flexibility to explore the different features' relative importance in cold question routing. For example, information of asker and question tags can be iteratively introduced to our model to explore their relative importance as follows:

\begin{itemize}[leftmargin=0.6cm]
\item {\em Seq-A}: explore the importance of question asker by using triples of $\langle \mathcal{Q},\mathcal{U},\mathcal{A} \rangle$ on routing cold questions asked by existing askers. Other settings are as same as {\em Seq-TA}.
\item {\em Seq-T}: explore the importance of question tags by using triples of $\langle \mathcal{Q},\mathcal{U},\mathcal{T} \rangle$ on routing cold questions either from existing askers or new askers. Other settings are as same as {\em Seq-TA}.
\item {\em Seq-Un}: explore the importance of question asker and tags by using tuples of $\langle \mathcal{Q},\mathcal{U} \rangle$ on routing cold questions. Other settings are as same as {\em Seq-TA}.
\end{itemize}

Regarding the regression module, we examined two types of SVM based regressors implemented by scikit-learn. One is a Epsilon-Support Vector Regression (SVM)\footnote{\url{http://scikit-learn.org/stable/modules/generated/sklearn.svm.SVR.html}} with the polynomial kernel (degree is set as $2$). Another is the LinearSVR\footnote{\url{http://scikit-learn.org/stable/modules/generated/sklearn.svm.LinearSVR.html}} with the kernel type as a linear function. A neural network based regressor which is implemented based on Keras \footnote{\url{https://keras.io}} has also been taken into examination. 

\subsection{The End-2-End Module for Cold Question Routing}

The end-2-end module consists of two layers: embedding layer and regression layer. The embedding layer is built based on the Graph Convolutional Networks (GCNs)~\cite{kipf2016semi}. The core idea behind GCNs is to learn how to iteratively aggregate feature information from local graph neighborhoods using neural networks~\cite{Ying2018GCN}. A single “convolution” operation transforms and aggregates feature information from a node’s one-hop graph neighborhood, and by stacking multiple such convolutions information can be propagated across far reaches of a graph~\cite{Ying2018GCN}. The embedding layer is followed by a multilayer perceptron (MLP) to do the regression. In our experiments, we use $2$ layers in GCNs and $3$ hidden layers in MLP. We evaluate the mean squared error for all the training examples. 

\subsection{Experts Finding for Cold Questions}

The final step of both {\em Seq} and {\em EndCold} is to identify the best answerers for cold questions, which works as follows:

\begin{enumerate}
    \item Given a cold question $q^*$ and a set of potential answerers $C_q^*$, predict each candidate $u$'s voting score for $q^*$, where $u \in C_q^*$
    \item Select the user who achieves the highest voting score as the best answerer for $q^*$
\end{enumerate}

Given a newly posted question $q^*$ asked by an asker $a$ using tags $t$ and a potential answerer $u$, the prediction function $f_{Seq}$ in {\em Seq} can treat the new question as a missing value by $f_{Seq}(\vec{0},\vec{u},\vec{a},\vec{t})$. The user $u$ who achieves the highest value of $f_{Seq}(\vec{0},\vec{u},\vec{a},\vec{t})$ will be selected as the best answerer for question $q^*$. It's possible that the newly posted question $q^*$ can be asked by a new asker who has no information to learn the corresponding embedding vector. Then the prediction function $f_{Seq}$ can treat the new asker as a missing value by $f_{Seq}(\vec{0},\vec{u},\vec{0},\vec{t})$. 

Instead of using missing value $\vec{0}$ to represent the feature vector of $q^*$ in {\em Seq}, {\em EndCold} can leverage GCNs to propagate information from existing tag nodes and asker nodes (if applicable) to $q^*$. $q^*$ can accumulate these information and get its corresponding embedding vector as $\vec{q^*}$. The regression function $f_{EndCold}$ in {\em EndCold} can predict the voting score $u$ can achieve on $q^*$ as $f_{EndCold}(\vec{q^*},\vec{u},\vec{a},\vec{t})$.

It is possible that the potential answerer $u$ is a newly registered user who has not provided any answer before in CQAs (user cold-start). In this scenario, the prediction function can be simplified as $f(\vec{q^*},\vec{0},\vec{a},\vec{t})$. All new registered users will receive the same predicted voting score for the same target question. More efforts will be spent to make accurate predictions for the user cold-start problem in our future work.

\section{Experiments and Analyses}
\label{sec:experiments}


We conducted our cold question routing experiments following the same settings as ColdRoute on $8$ large Stack Exchange sites~\cite{ColdRoute}. We compare our proposed models with state-of-the-art methods based on two popular evaluation criteria ({\bf Precision$@3$} ~\cite{Zhu2014,zhaoexpert2016,Fang2016QuestionAnswering,zhao2017community,ColdRoute,Sun2018ATP} and {\bf Accuracy} ~\cite{zhaoexpert2016,Fang2016QuestionAnswering,zhao2017community,ColdRoute,Sun2018ATP})\footnote{Other evaluation criteria such as Precision$@1$ and MRR~\cite{Zhu2014} used in ColdRoute~\cite{ColdRoute} follow the same performance pattern as Precision$@3$ and Accuracy. To save space, we don't report their results here.}. 


\subsection{Performance of Different Sequential Models}
\label{sec:sub_sec_cold_question_routing_embedding_methods}

In this section, we compare the routing performance of different sequential models with different embedding methods (such as {\em node2vec} and {\em ATP}) on various input graphs (such as directed competition graph used by {\em QDEE} and undirected CQA graph with question tags proposed in this work) with our proposed models. The key steps of each method in this section follow the same sequential framework as we discussed in Methodology section. The differences fall in how to build the input graph and get the embedding vector for each node in the corresponding input graph. The details of all comparison partners are as follows:


\begin{itemize}[leftmargin=0.4cm]
\item  {\bf  Seq-Dir-ATP}: The input graph of {\em Seq-Dir-ATP} is the same directed competition graph used in {\em QDEE}. Embeddings of nodes (questions and users) are generated by {\em ATP}. Given a newly posted question $q^*$, {\em Seq-Dir-ATP} applies Expertise Gain Assumption ({\em EGA})~\cite{QDEE2018} to use the source and target embedding of the question which has the highest difficulty level among all questions asked by the same asker to approximate $q^*$'s source and target embedding respectively. Given a candidate user $u$, the number of votes $u$ can achieve on $q^*$ can be predicted by feeding the concatenated source and target feature vectors of $q^*$ and $u$ to a regression model $f$, which can be represented as $f(\vec{q_s^*}, \vec{q_t^*},\vec{u_s}, \vec{u_t})$, where $\vec{._s}$ and $\vec{._t}$ represent corresponding node's source and target feature vector respectively. The dimension  of both source and target feature vector is tuned to set as $128$. Hence the dimension of concatenated feature for $f$ is $512$. 

\item {\bf Seq-Dir}: It uses the same directed competition graph as {\em Seq-Dir-ATP}. However, {\em Seq-Dir} leverages node2vec~\cite{grovernode2vec} to generate embeddings for question and user nodes. Like {\em Seq-Dir-ATP}, {\em EGA} is applied to generate embeddings for newly posted questions. Feature vector representations of $q^*$ and $u$ are concatenated as inputs of the regression model. Parameters are tuned to set as $p=q=1, d = 128$\footnote{\url{https://github.com/snap-stanford/snap/tree/master/examples/node2vec}}. Hence the dimension of concatenated feature for $f$ is $256$.
\end{itemize}

It's worth mentioning {\em Seq-Dir-ATP} and {\em Seq-Dir} cannot handle the scenario of cold questions asked by new askers since {\em EGA} can only be applied to questions asked by existing askers to infer the embedding vector of the newly posted question. Hence, we only show their corresponding performance on routing cold questions asked by existing askers. 
The experiment results are summarized in Figure~\ref{fig:ColdStart_SequentialModels}, where we can make the following observations:

\begin{figure*}[ht!]
        \centering
        \begin{subfigure}[b]{0.48\textwidth}
            \centering
           \includegraphics[width=\textwidth]{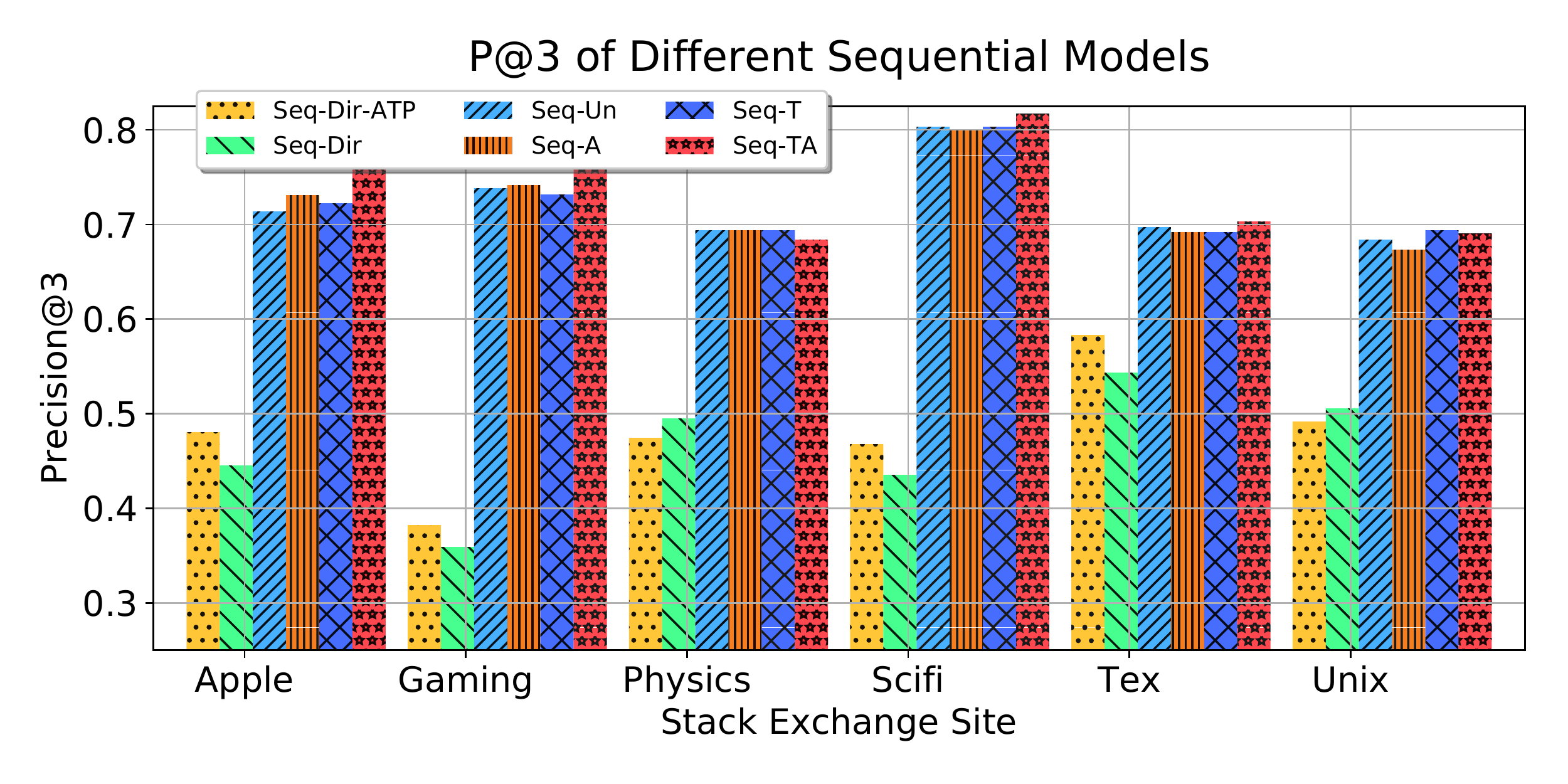}
            \caption[Precision$@$3]%
            {{\small Precision$@$3}} 
            \label{fig:mean and std of net14}
        \end{subfigure}
        \hfill
        \begin{subfigure}[b]{0.48\textwidth}  
            \centering 
           \includegraphics[width=\textwidth]{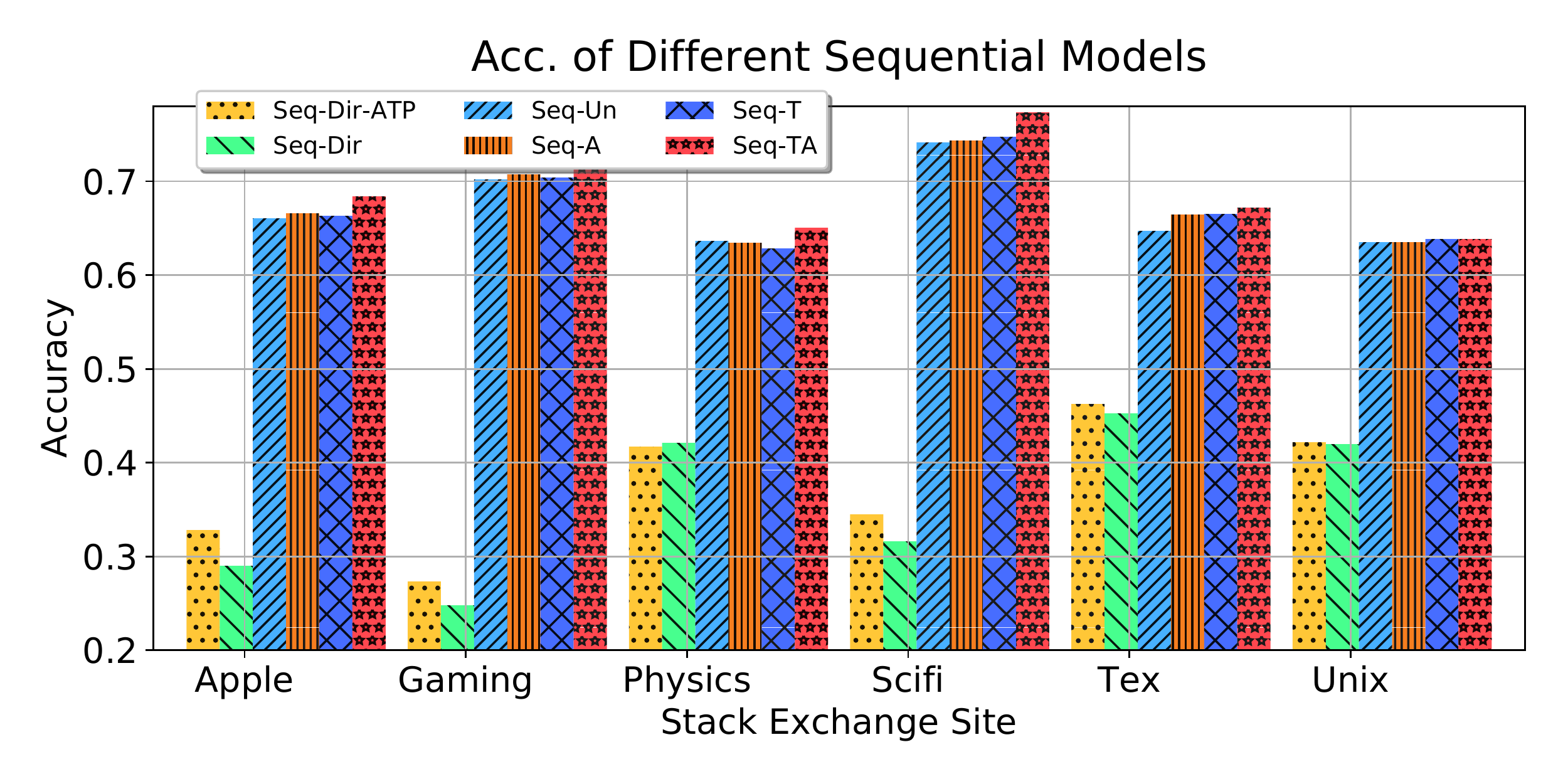}
            \caption[]%
            {{\small Accuracy}}    
            \label{fig:mean and std of net24}
        \end{subfigure}
        \caption[]
        {Performance of Different Sequential Models on Routing Cold Questions Posted by Existing Askers} 
        \label{fig:ColdStart_SequentialModels}
        \vspace{-0.1in}
    \end{figure*}

\begin{itemize}[leftmargin=0.4cm]
    \item The best overall performance is achieved by {\em Seq-AT}, which concatenates the embedding of all four nodes in the $(\vec{q},\vec{u},\vec{a},\vec{t})$ quadruple as input features and feeds them to the regression model. The fact that {\em Seq-AT} consistently outperforms its feature-ablated variants {\em Seq-A}, {\em Seq-T} and {\em Seq-Un} gives us some guideline that it is necessary to incorporate both user past activities (asking and answering) and textual information when designing question routing algorithms.
    \item For all the CQA sites and under all metrics, the four methods using the undirected CQA graph with question tags as input perform significantly better than {\em Seq-Dir} and {\em Seq-Dir-ATP}) which use the direct competition graph. It demonstrates the advantage of using our proposed undirected CQA Graph with question tags in comparison with using the directed competition graph leveraged by {\em QDEE}~\cite{QDEE2018} and {\em ATP}~\cite{Sun2018ATP} on the task of cold question routing. 
    \item Note that despite {\em Seq-Un} does not explicitly use the embedding of question tag nodes, it still significantly outperforms {\em Seq-Dir}. This is because the graph embedding algorithm (node2vec~\cite{grovernode2vec}) in our experiments) allows neighboring nodes to share information, so the question node embeddings are regulated by their assigned tags as well.
\end{itemize}

\begin{table*}[ht!]
\centering
\caption{Performance on newly posted questions asked by existing askers in $8$ different Stack Exchange sites} 
\resizebox{.99\textwidth}{!}{
\begin{tabular}{@{}c|c|cccccccc@{}} \toprule
 & \boldmath\textbf{Site} & \boldmath\textbf{Apple} & \boldmath\textbf{AskUbuntu} & \boldmath\textbf{Gaming} & \boldmath\textbf{Physics} & \boldmath\textbf{Scifi} & \boldmath\textbf{Serverfault} & \boldmath\textbf{Tex} & \boldmath\textbf{Unix}\\ \midrule
 
 
 & Doc2Vec & 0.3889 & 0.4133 & 0.2971 & 0.3571 & 0.276 & 0.3979 & 0.5029 & 0.4175\\
 & \cellcolor[rgb]{ .851,  .851,  .851}ColdRoute-T & \cellcolor[rgb]{ .851,  .851,  .851}0.6581 & \cellcolor[rgb]{ .851,  .851,  .851}0.6274 & \cellcolor[rgb]{ .851,  .851,  .851}0.7796 & \cellcolor[rgb]{ .851,  .851,  .851}\boldmath\textbf{0.7194} & \cellcolor[rgb]{ .851,  .851,  .851}0.7742 & \cellcolor[rgb]{ .851,  .851,  .851}0.6074 & \cellcolor[rgb]{ .851,  .851,  .851}0.6343 & \cellcolor[rgb]{ .851,  .851,  .851}0.6869\\

& ColdRoute-A & 0.6026 & 0.5889 & 0.7157 & 0.6582 & 0.6846 & 0.5799 & 0.5657 & 0.5690 \\
\boldmath\textbf{P@3} & \cellcolor[rgb]{ .851,  .851,  .851}ColdRoute-TA & \cellcolor[rgb]{ .851,  .851,  .851}0.5641 & \cellcolor[rgb]{ .851,  .851,  .851}0.5717 & \cellcolor[rgb]{ .851,  .851,  .851}0.6805 & \cellcolor[rgb]{ .851,  .851,  .851}0.6939 & \cellcolor[rgb]{ .851,  .851,  .851}0.7599 & \cellcolor[rgb]{ .851,  .851,  .851}0.5778 & \cellcolor[rgb]{ .851,  .851,  .851}0.6114 & \cellcolor[rgb]{ .851,  .851,  .851}0.6667 \\

 & \boldmath\textbf{Seq} (pointwise) & 0.7607 & 0.668 & 0.8019 & 0.6837 & 0.8172 & 0.6 & \boldmath\textbf{0.7029} & 0.6902\\ 
 & \cellcolor[rgb]{ .851,  .851,  .851}\boldmath\textbf{Seq} (pairwise) & \cellcolor[rgb]{ .851,  .851,  .851}\boldmath\textbf{0.7692} & \cellcolor[rgb]{ .851,  .851,  .851}0.7345	 & \cellcolor[rgb]{ .851,  .851,  .851}0.8147 & \cellcolor[rgb]{ .851,  .851,  .851}0.6786 & \cellcolor[rgb]{ .851,  .851,  .851}\boldmath\textbf{0.8208} & \cellcolor[rgb]{ .851,  .851,  .851}0.6117 & \cellcolor[rgb]{ .851,  .851,  .851}0.6286 & \cellcolor[rgb]{ .851,  .851,  .851}\boldmath\textbf{0.7003} \\
 & \boldmath\textbf{EndCold} & 0.7607 & \boldmath\textbf{0.743} & \boldmath\textbf{0.8339} & 0.7041 & \boldmath\textbf{0.8208} & \boldmath\textbf{0.6222} & 0.6914 & 0.6902 \\
 
 \hline
 & Doc2Vec & 0.3076 & 0.4333 & 0.3315 & 0.3641 & 0.3097 & 0.416 & 0.4897 & 0.4044\\
 & \cellcolor[rgb]{ .851,  .851,  .851}ColdRoute-T & \cellcolor[rgb]{ .851,  .851,  .851}0.6324 & \cellcolor[rgb]{ .851,  .851,  .851}0.6054 & \cellcolor[rgb]{ .851,  .851,  .851}0.7387 & \cellcolor[rgb]{ .851,  .851,  .851}0.6354 & \cellcolor[rgb]{ .851,  .851,  .851}0.7369 & \cellcolor[rgb]{ .851,  .851,  .851}0.5807 & \cellcolor[rgb]{ .851,  .851,  .851}0.5802 & \cellcolor[rgb]{ .851,  .851,  .851}0.6404\\

& ColdRoute-A & 0.5822 & 0.5814 & 0.6710 & 0.6159 & 0.6690 & 0.5655 & 0.5498 & 0.5422 \\
\boldmath\textbf{Acc} & \cellcolor[rgb]{ .851,  .851,  .851}ColdRoute-TA & \cellcolor[rgb]{ .851,  .851,  .851}0.5573 & \cellcolor[rgb]{ .851,  .851,  .851}0.5671 & \cellcolor[rgb]{ .851,  .851,  .851}0.6596 & \cellcolor[rgb]{ .851,  .851,  .851}{0.6381} & \cellcolor[rgb]{ .851,  .851,  .851}0.7174 & \cellcolor[rgb]{ .851,  .851,  .851}0.5579 & \cellcolor[rgb]{ .851,  .851,  .851}0.5727 & \cellcolor[rgb]{ .851,  .851,  .851}0.6174 \\

 & \boldmath\textbf{Seq} (pointwise) & 0.6836 & 0.6572 & 0.759 & 0.65 & 0.7734 & 0.5917 & \boldmath\textbf{0.6719} & 0.6381\\
 & \cellcolor[rgb]{ .851,  .851,  .851}\boldmath\textbf{Seq} (pairwise) & \cellcolor[rgb]{ .851,  .851,  .851}\boldmath\textbf{0.7079} & \cellcolor[rgb]{ .851,  .851,  .851}0.7092 & \cellcolor[rgb]{ .851,  .851,  .851}0.7656 & \cellcolor[rgb]{ .851,  .851,  .851}0.6404 & \cellcolor[rgb]{ .851,  .851,  .851}0.7619 & \cellcolor[rgb]{ .851,  .851,  .851}0.6005 & \cellcolor[rgb]{ .851,  .851,  .851}0.5888 & \cellcolor[rgb]{ .851,  .851,  .851}\boldmath\textbf{0.648} \\
& \boldmath\textbf{EndCold} & 0.7022 & \boldmath\textbf{0.7189} & \boldmath\textbf{0.778} & \boldmath\textbf{0.6733} & \boldmath\textbf{0.7787} & \boldmath\textbf{0.607} & 0.6495 & 0.6401 \\
 \bottomrule
\end{tabular}
 }
\label{tab:node2vec_quad_cold_routing_ne}
\end{table*}

\begin{table*}
\centering
\caption{Performance on newly posted questions asked by new askers in $8$ different Stack Exchange sites} 
\resizebox{.99\textwidth}{!}{
\begin{tabular}{c|c|cccccccc}
 & \boldmath\textbf{Site} & \boldmath\textbf{Apple} & \boldmath\textbf{AskUbuntu} & \boldmath\textbf{Gaming} & \boldmath\textbf{Physics} & \boldmath\textbf{Scifi} & \boldmath\textbf{Serverfault} & \boldmath\textbf{Tex} & \boldmath\textbf{Unix}\\  \midrule
 
 & Doc2Vec & 0.384 & 0.4096 & 0.3563 & 0.3493 & 0.2547 & 0.4 & 0.4848 & 0.4297\\
 & \cellcolor[rgb]{ .851,  .851,  .851}ColdRoute-T & \cellcolor[rgb]{ .851,  .851,  .851}0.5171 & \cellcolor[rgb]{ .851,  .851,  .851}0.61 & \cellcolor[rgb]{ .851,  .851,  .851}0.7 & \cellcolor[rgb]{ .851,  .851,  .851}\boldmath\textbf{0.7249} & \cellcolor[rgb]{ .851,  .851,  .851}0.7081 & \cellcolor[rgb]{ .851,  .851,  .851}0.6074 & \cellcolor[rgb]{ .851,  .851,  .851}0.6894 & \cellcolor[rgb]{ .851,  .851,  .851}0.6525 \\


  & ColdRoute-TA & {0.5589} & 0.6013 & 0.6688 & 0.7205 & 0.6957 & 0.5617 & 0.6439 & 0.6419 \\

 \boldmath\textbf{P@3} & \cellcolor[rgb]{ .851,  .851,  .851}\boldmath\textbf{Seq} (pointwise) & \cellcolor[rgb]{ .851,  .851,  .851}0.6502 & \cellcolor[rgb]{ .851,  .851,  .851}0.6863 & \cellcolor[rgb]{ .851,  .851,  .851}0.7625 & \cellcolor[rgb]{ .851,  .851,  .851}0.7205 & \cellcolor[rgb]{ .851,  .851,  .851}\boldmath\textbf{0.764} & \cellcolor[rgb]{ .851,  .851,  .851}0.6267 & \cellcolor[rgb]{ .851,  .851,  .851}0.7121 & \cellcolor[rgb]{ .851,  .851,  .851}0.6499\\ 
 &  \boldmath\textbf{Seq} (pairwise) &  	0.6616 & 	0.6885 & 	0.7625 & 	0.6681 & 	0.7453 & 	0.61 & 	0.7045 & 	0.671 \\
 &   \cellcolor[rgb]{ .851,  .851,  .851}\boldmath\textbf{EndCold}	 &  \cellcolor[rgb]{ .851,  .851,  .851}\boldmath\textbf{0.7034} & 	\cellcolor[rgb]{ .851,  .851,  .851}\boldmath\textbf{0.7102} & 	\cellcolor[rgb]{ .851,  .851,  .851}\boldmath\textbf{0.8} & \cellcolor[rgb]{ .851,  .851,  .851}\boldmath\textbf{0.7249} & 	\cellcolor[rgb]{ .851,  .851,  .851}0.7516	 & \cellcolor[rgb]{ .851,  .851,  .851}\boldmath\textbf{0.63}	 & \cellcolor[rgb]{ .851,  .851,  .851}\boldmath\textbf{0.7349} & 	\cellcolor[rgb]{ .851,  .851,  .851}\boldmath\textbf{0.6976}\\
 \hline
 & Doc2Vec & 0.4 & 0.4138 & 0.3779 & 0.3803 & 0.3044 & 0.4233 & 0.4308 & 0.4274\\
 & \cellcolor[rgb]{ .851,  .851,  .851}ColdRoute-T & \cellcolor[rgb]{ .851,  .851,  .851}0.5247 & \cellcolor[rgb]{ .851,  .851,  .851}0.5809 & \cellcolor[rgb]{ .851,  .851,  .851}0.6591 & \cellcolor[rgb]{ .851,  .851,  .851}0.6838 & \cellcolor[rgb]{ .851,  .851,  .851}0.6841 & \cellcolor[rgb]{ .851,  .851,  .851}0.5807 & \cellcolor[rgb]{ .851,  .851,  .851}0.5747 & \cellcolor[rgb]{ .851,  .851,  .851}0.6034\\


& ColdRoute-TA & {0.5555} & 0.5797 & 0.6327 & 0.6810 & 0.6761 & 0.5460 & 0.5700 & 0.5995 \\ 

\boldmath\textbf{Acc} & \cellcolor[rgb]{ .851,  .851,  .851}\boldmath\textbf{Seq} (pointwise) & \cellcolor[rgb]{ .851,  .851,  .851}0.6179 & \cellcolor[rgb]{ .851,  .851,  .851}0.6219 & \cellcolor[rgb]{ .851,  .851,  .851}0.7123 & \cellcolor[rgb]{ .851,  .851,  .851}0.6853 & \cellcolor[rgb]{ .851,  .851,  .851}\boldmath\textbf{0.7356} & \cellcolor[rgb]{ .851,  .851,  .851}\boldmath\textbf{0.5955} & \cellcolor[rgb]{ .851,  .851,  .851}0.6069 & \cellcolor[rgb]{ .851,  .851,  .851}0.6163 \\ 
 & \boldmath\textbf{Seq} (pairwise) & 0.6326 & 0.6455 & 0.708 & 0.6604 & 0.7354 & 0.578 &	0.587 & 0.6261  \\
 &  \cellcolor[rgb]{ .851,  .851,  .851}\boldmath\textbf{EndCold} & \cellcolor[rgb]{ .851,  .851,  .851}\boldmath\textbf{0.6743} & \cellcolor[rgb]{ .851,  .851,  .851}\boldmath\textbf{0.6881} & \cellcolor[rgb]{ .851,  .851,  .851}\boldmath\textbf{0.7447} & \cellcolor[rgb]{ .851,  .851,  .851}\boldmath\textbf{0.6887} & \cellcolor[rgb]{ .851,  .851,  .851}0.731 & 	\cellcolor[rgb]{ .851,  .851,  .851}0.5915 & \cellcolor[rgb]{ .851,  .851,  .851}\boldmath\textbf{0.6227} & \cellcolor[rgb]{ .851,  .851,  .851}\boldmath\textbf{0.6393}\\ \bottomrule
 \end{tabular}
 }
\label{tab:node2vec_quad_cold_routing_nn}
\end{table*}
\subsection{Pointwise v.s. Pairwise}

 In this section, We would like to compare the performance difference between pointwise and pairwise based sequential models on cold question routing in CQAs, with hoping that the corresponding results can guide our future design of the question routing system in CQAs. The difference between pointwise and pairwise based sequential models are listed as follows:

\begin{itemize}[leftmargin=*]
    \item {\bf Pointwise}: its goal is to learn a regression function $f: \langle q,u,a,t \rangle \rightarrow \mathbb{R}$. The user $u \in \mathcal{U}$ who achieves the highest value of $f(q,u,a,t)$ will be selected as the best answerer for question $q$. 
    \item {\bf Pairwise}: We introduce the relative quality rank to model users' past answering activities in CQAs, which is in the form of $(q, a, t, u_+, u_{\_})$, meaning that the answer provided by answerer $u_+$, obtains more voting scores than the answer provided by the answerer $u_{\_}$ for question $q$ asked by asker $a$ with tags set $t$. Let $\mathcal{C} = \{ (q, a, t, u_+, u_{\_}) \}$ denote the set of ranking pair constraints derived from the community votes. More formally, we aim to learn a ranking function $f$ that for any $(q, a, t, u_+, u_{\_}) \in \mathcal{C}$, the inequality holds: $f(q, a, t, u_+) > f(q, a, t, u_{\_}) + \epsilon$. The user $u_+ \in \mathcal{U}$ who satisfies the most number of above constraints will be selected as the best answerer for question $q$. Given the representations of question and answerers, and the relative quality rank in CQA sites, the loss function $\mathcal{L}_r$ in Pairwise is designed as follows:

        \begin{multline}
        \mathcal{L}_r = \sum_{(q, a, t, u_+, u_{\_}) \in \mathcal{C}} 
        \\
        = \max(0, \epsilon + f^{\_}(q, a, t, u_{\_}) - f^{+}(q, a, t, u_+)))
        \label{eq:loss_func}
        \end{multline}

    where $f^{+}(\cdot)$ denotes the voting score achieved by high-quality answerers and $f^{\_}(\cdot)$ represents the voting score of low-quality answerers for question routing. The hyper-parameter $\epsilon$ ($\epsilon > 0$) controls the margin in the loss function and $\mathcal{C}$ is the set of pairwise relative ranking constraints.
\end{itemize}

In our experiments, the pairwise based model {\em Seq (pairwise) } leveraged $SVM^{rank}$~\cite{ranksvm} to learn the partial order of two potential answerers given a newly posted question. As shown in Table~\ref{tab:node2vec_quad_cold_routing_ne} (cold questions asked by existing askers) and ~\ref{tab:node2vec_quad_cold_routing_nn} (cold questions asked by new askers), {\em Seq (pairwise)} does not show significant improvement over the pointwise model {\em Seq (pointwise)} consistently. The reason is that {\em Seq (pairwise)} suffers from the data sparseness problem. Table~\ref{tab:average_number_answers_per_question} shows the statistical information of $8$ Stack Exchange sites used in our experiments. The average number of answers per question among the $8$ Stack Exchange sites is $1.79$, and $57.19\%$ of questions have only answer, which indicate that {\em Seq (pairwise)} cannot extract enough valid question-answerer pairs for training. With CQAs growing and information of answers becoming dense, {\em Seq (pairwise)} can have more valid relative ranking constraints in $\mathcal{C}$ to train and then be more robust and effective.

\begin{table*}[]
\centering
\caption{Statistics of Stack Exchange Sites on Average Number of Answers per Question} 
\resizebox{.99\textwidth}{!}{
\begin{tabular}{c|cccccccc}
   & \textbf{Apple} & \textbf{AskUbuntu} & \textbf{Gaming} & \textbf{Physics} & \textbf{Scifi} & \textbf{Serverfault} & \textbf{Tex} & \textbf{Unix}\\ \hline
\textbf{Avg \# answers per question} & 1.8122 & 1.6391 & 1.7796 & 1.7823 & 2.1405 & 1.8521 & 1.5371 & 1.7917\\ \hline 
\textbf{\# questions with only one answer (percentage)} & 0.5803 & 0.6339 & 0.5723 & 0.5516 & 0.4876 & 0.5463 & 0.6507 & 0.5524 \\ \hline 
\end{tabular}}
\label{tab:average_number_answers_per_question}
\end{table*}

\subsection{Sequential Model v.s. End-2-End Model}

In this section, we aim to compare the difference between our sequential model and end-2-end model. As shown in Table~\ref{tab:node2vec_quad_cold_routing_ne} and~\ref{tab:node2vec_quad_cold_routing_nn}, the overall performance of end-2-end model {\em EndCold} is better than the performance of sequential models (both pointwise and pairwise). For example, {\em EndCold} improves upon the evaluation metrics (Accuracy and Precision$@3$) over {\em Seq} (pointwise) by $3.37\%$ and $3.33\%$ respectively on routing cold questions asked by new askers. Above results demonstrate the effectiveness of our end-2-end framework in comparing with the sequential models. 

\subsection{Comparison with the State-of-the-art on Cold Question Routing}

In this section, we compared our proposed model with {\em ColdRoute}~\cite{ColdRoute}, a state-of-the-art model on routing cold questions posted by both new and existing askers to suitable experts. 
{\em ColdRoute} encodes users' past asking and answering activities by using one-hot encoding, which can cause sparsity problem. 
Instead of using one-hot encoding and then suffering from the sparsity problem, our proposed model leverages the dense embedding vectors of question, asker, answerer and question tags (average) to build the input feature vector for training. It's worth mentioning that {\em ColdRoute} fails to take the higher-order interactions among questions, askers, answerers, and question tags into consideration due to the high complexity. 
{\em ColdRoute-A}, {\em ColdRoute-T}, and {\em ColdRoute-TA} are three variants of {\em ColdRoute} and they were designed to explore the importance of asker information, question tags, and both respectively.

We also compare our model with some semantic matching based approaches ~\cite{Srba2016,Li2010RQA,Li2011QuestionRouting,CQAWord2Vec,Yang2014EUE,Szpektor2013,Yang2013CQArank,ColdRoute}.
These matching models have demonstrated their power on finding suitable experts recently~\cite{Srba2016}. 
Due to space limitation\footnote{Other semantic matching models such as Bag of Words~\cite{Figueroa2013LRE-BOW,IJCAIZhou2013-BOW} and  Latent Dirichlet Allocation~\cite{Guo2008-LDA,Ji2012LDA}, and Word2Vec~\cite{Mikolov2013} follow the same pattern as Doc2Vec}, we only show the performance of {\bf Doc2Vec} ~\cite{doc2vec,CQAWord2Vec} 
The Doc2Vec used in our paper is implemented by gensim \footnote{\url{https://radimrehurek.com/gensim/models/doc2vec.html}}. The dimension of the feature vector is tuned to set as $80$. 

As shown in Table~\ref{tab:node2vec_quad_cold_routing_ne} and~\ref{tab:node2vec_quad_cold_routing_nn}, our proposed models perform better than {\em ColdRoute} ({\em ColdRoute-A}, {\em ColdRoute-T}, and {\em ColdRoute-TA}) and {\em Doc2Vec} on routing cold questions asked by both existing askers and new askers, which demonstrates the effectiveness of our proposed framework. 

\section{Conclusion and Future Work}
\label{sec:conclusion}

In this paper, we proposed an end-to-end framework for cold question routing in CQAs. The input of the framework is an undirected heterogeneous CQA graph with encodes users' past asking and answering activities and textual information of questions. Our proposed model can leverage the higher-order graph structure and content information to embed nodes in the input graph, and do the question routing simultaneously. Extensive experiments show that our model performs better than the state-of-the-art models on routing cold questions (asked by both existing askers and new askers).



As a future work, we plan to test our models on other CQAs with different settings (such as having more dense askers and answerers). In order to increase the expertise of the entire community, we plan to address the problem of routing newly posted questions (item cold-start) to newly registered users (user cold-start) in CQAs.

 {\bf Acknowledgments} This work is supported by NSF grants EAR-1520870, CCF-1645599, CCF-1629548, IIS-1550302, and CNS-1513120, and a grant from the Ohio Supercomputer Center (PAS0166). All content represents the opinion of the authors, which is not necessarily shared or endorsed by their sponsors.


\bibliographystyle{aaai}
\bibliography{sigproc}

\end{document}